\documentclass[doublecol]{epl2}
\usepackage{graphicx,amsmath,bbm,mathrsfs,amssymb,pstricks,psfrag}

\newcommand{\dket}[1]{\vert  #1  \rangle\rangle}
\newcommand{\dbra}[1]{\langle\langle #1 \vert}
\newcommand{\bra}[1]{\langle #1 \vert}
\newcommand{\ket}[1]{\vert #1 \rangle}

\newcommand{\Tr}[2]{\textup{Tr}_{#1} \left[#2\right]}

\newcommand{\R}{{\scriptstyle R}}
\title{Conditional measurements on multimode pairwise entangled
states from spontaneous parametric downconversion}
\author{Alessia Allevi$^{1}$, Alessandra Andreoni$^{2,1}$, Federica A.
Beduini$^{4}$, Maria
Bondani$^{3,1}$, Marco G. Genoni$^{5,4}$, Stefano Olivares$^{5,4}$ \and
Matteo G. A. Paris$^{4,5}$}
\shortauthor{A. Allevi et al.}
\institute{
${}^{1}$CNISM UdR Como, I-22100 Como, Italia\\
${}^{2}$Dipartimento di Fisica e Matematica, Universit\`a degli Studi
dell'Insubria, I-22100 Como, Italia\\
${}^{3}$Istituto di Fotonica e Nanotecnologie - CNR-IFN, I-22100, Como, Italia\\
${}^{4}$Dipartimento di Fisica dell'Universit\`a degli Studi di Milano, I-20133 Milano, Italia. \\
${}^{5}$CNISM, UdR Milano Universit\`a, I-20133 Milano, Italia.}
\abstract{We address the intrinsic multimode nature of the quantum state
of light obtained by pulsed spontaneous parametric downconversion and
develop a theoretical model based only on experimentally accessible
quantities.  We exploit the pairwise entanglement as a resource for
conditional multimode measurements and derive closed formulas for the
detection probability and the density matrix of the conditional states.
We present a set of experiments performed to validate our model in
different conditions that are in excellent agreement with experimental
data. Finally, we evaluate nonGaussianity of the conditional states obtained
from our source with the aim of discussing the
effects of the different experimental parameters on the efficacy of this
type of conditional state preparation.}
\date{\today}
\pacs{03.67.Bg}{Entanglement production and manipulation}
\pacs{42.50.Dv}{Quantum state engineering and measurements}
\pacs{03.65.Wj}{State reconstruction, quantum tomography}
\begin{document}
\maketitle
Nonclassical states of the radiation field represent a resource for quantum information
and communication and much attention has been devoted to their generation schemes. As a
matter of fact, beside squeezing, nonclassical effects are generally observed in
connection with nonGaussian states of light, and this usually implies the presence of
fluctuating parameters \cite{ngOPO1,ngOPO2} or of nonlinearities higher than the second
order are involved (\emph{e.g.} the Kerr effect
\cite{kerr:PRL:01,kerr:PRA:06,kerr:NJP:08}) in the generation scheme. On the other hand,
the reduction postulate provides an alternative mechanism to achieve effective nonlinear
dynamics. In fact, if a measurement is performed on a portion of a composite entangled
system, the other component is conditionally reduced according to the outcome of the
measurement. The resulting dynamics may be highly nonlinear, and may produce quantum
states that cannot be generated by currently achievable nonlinear processes \cite{eng}.
The efficiency of the process, \emph{i.e.} the rate of success in getting a certain
state, is equal to the probability of obtaining a certain outcome from the measurement
and it may be higher than nonlinear efficiency, thus making conditional schemes possibly
convenient even when a corresponding Hamiltonian process exists.
\par
The nonlinear dynamics induced by conditional measurements has been analyzed for a large
variety of schemes
\cite{eng,dar,wel,mar:PRA:10,coc,oli,oli1,pgb,cat1,cat2,cat3,ff,zlc,uni,
opq,cla,sab,pla,bno,koz,kim}, including photon addition and subtraction schemes
\cite{wel,coc,oli,mar:PRA:10,all10}, optical state truncation of coherent states \cite{pgb},
generation of cat-like states \cite{cat1,cat2,cat3}, state filtering by active cavities
\cite{ff,zlc}, synthesis of arbitrary unitary operators \cite{uni} and generation of
optical qubit by conditional interferometry \cite{opq}. More recently, nonGaussianity of
states and operations has been recognized as a relevant resource for a series of tasks
including entanglement distillation \cite{eisert:entdist,fiur:entdist,ngnp1,ngnp2} and
improvements in both teleportation \cite{wel,coc,oli,inv:PRA:05}, cloning~\cite{c1} and
storage \cite{cs07}. Conditional state generation has been achieved in the low energy
regime~\cite{ourjo:entdist,taka09,gerrits10} by using single-photon detectors, and the
question arises whether analogue schemes may be used also in the mesoscopic domain
\cite{mas06}.
\par
In this paper we address multimode conditional measurements and demonstrate a novel
bright source of nonclassical states \cite{ps10} based on (i) pulsed multimode
spontaneous parametric down-conversion (PDC)
\cite{jd05,Was06,ssh07,Lvo07,Was08,sil09a,sil09b,sil10}, which
produces entangled states with a mesoscopic number of photons and (ii) a conditional
intensity measurement performed by photoemissive detectors, called hybrid photodetectors,
that are able to partially resolve the number of detected photons~\cite{bjmo09}.  We
develop a theoretical model based only on experimentally accessible quantities and derive
closed formulas for both the detection probabilities and the conditional states.  We find
an excellent agreement with the experimental data and succeed in evaluating the amount of
nonGaussianity of the conditional states despite the multimode character of the entangled
state.
\par
\begin{figure}[ht]
\centerline{\includegraphics[width=.27\columnwidth,angle=270]{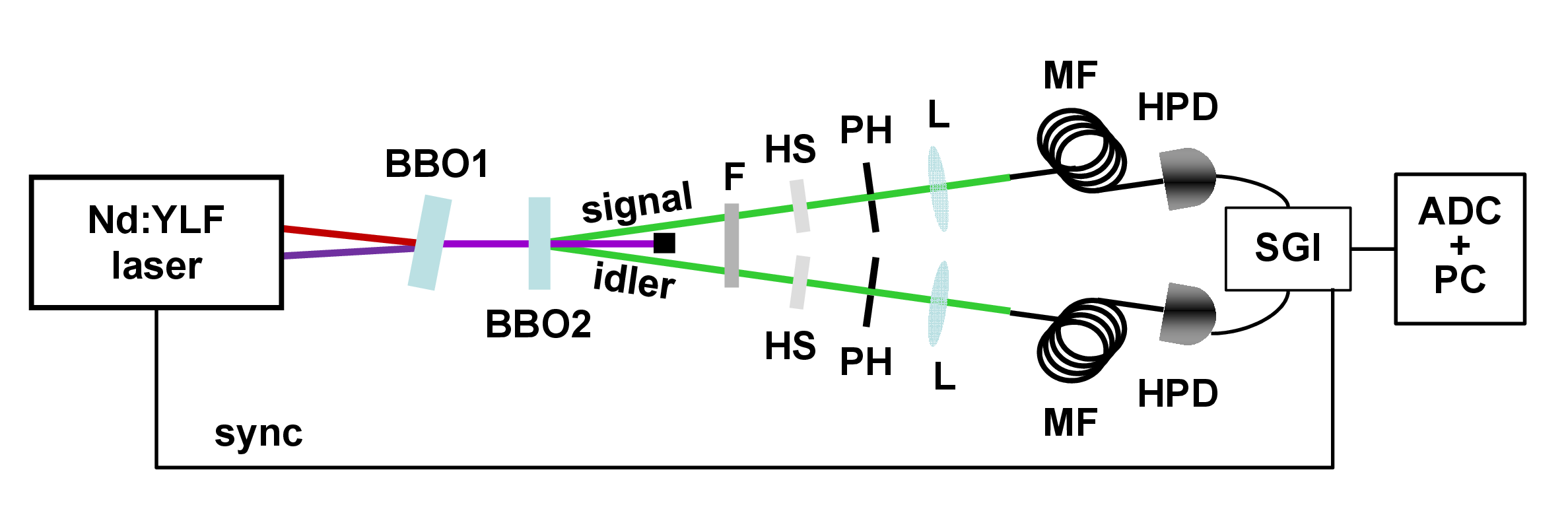}}
\caption{(Color online) Schematic diagram of the experimental setup.
BBO1 and BBO2: nonlinear crystals; F: cut-off filter; HS: harmonic
separators; PH: pin-hole apertures; L: lenses; MF: multimode optical fibers;
HPD: amplified hybrid photodetectors; SGI: synchronous gated integrator; PC:
digitizing PC board.} \label{setup}
\end{figure}

The pair of intense correlated beams obtained by pulsed PDC represents a convenient
system for state preparation by conditional measurements. In this case, the state
outgoing the crystal is intrinsically multimode because of the pulsed nature of the pump
and the properties of the nonlinear interaction~\cite{pale04,ssh07}, whereas correlations
are provided by the pairwise entanglement induced by spontaneous PDC. If we assume that
the output energy is equally distributed among the $\mu$ modes of each beam, then the
overall multimode state produced by pulsed PDC can be written as a tensor product of
$\mu$ identical twin-beam states, \emph{i.e.},
\begin{align}
\boldsymbol{R}&= \bigotimes_{k=1}^\mu
\dket{\lambda}_k{}_k \dbra{\lambda} \notag \\
\dket{\lambda}&=\sqrt{1-\lambda^2}\sum_n
\lambda^n |n\rangle\otimes|n\rangle \notag
\end{align}
with $\lambda^2=N/(\mu+N)$, $N$ being the mean total number of photons in either of the
two beams.  In our scheme, which is sketched in Fig.~\ref{setup}, conditional preparation
is obtained when one of the two beams undergoes a photon counting process. If we assume
that the detector efficiency $\eta$ is the same for each of the $\mu$ modes, the
probability operator-valued measure (POVM)$\{\boldsymbol{\Pi}_m\}$ describing the
detection of $m$ photoelectrons may be written as $$\boldsymbol{\Pi}_m =
\sum_{\boldsymbol q} \delta_{m\gamma} \bigotimes_{j=1}^{\mu}{\Pi}_{q_j}\:,$$ where
${\boldsymbol q}= \{ q_1, \dots, q_\mu \}$, $\gamma = \sum_{k=1}^\mu q_k$, $\delta_{hk}$
is the Kronecker delta, and
$${\Pi}_q  = \eta^q \sum_{k=q}^\infty (1-\eta)^{k-q} \binom{k}{q}
\ket{k}\bra{k}$$ denotes single-mode photon counting POVM.
 The joint probability distribution of
photoelectrons is given by $p_{12}(s,t) = \Tr{12}{\boldsymbol{R}
\,\boldsymbol{\Pi}_{s} \otimes \boldsymbol{\Pi}_{t}}$, that, after some
algebra, reads
\begin{align}\label{uno}
p_{12}&(s,t) = \left(\frac{\mu
\eta}{M+\mu \eta}\right)^\mu
\left(\frac{\eta}{1-\eta}\right)^{s+t} \notag\\
&\times \sum_{l=\max(s,t)}^\infty
\left[\frac{M(1-\eta)^2}{M+\mu \eta} \right]^l \binom{l + \mu
-1}{l} \binom{l}{s} \binom{l}{t}\;,
\end{align}
where $M=\frac12 \hbox{Tr}_{12}[\boldsymbol{R}\, \sum_{s} s\, \boldsymbol{\Pi}_{s}\otimes
\sum_{t} t\, \boldsymbol{\Pi}_{t} ]=\eta N$ is the total mean number of photoelectrons
measured on each of the two beams.  Notice that (\ref{uno}) only contains quantities that
can be experimentally accessed by direct detection.
\par
When one beam is detected, say the idler, and $t$ photoelectrons are obtained in the
measurement, the corresponding conditional state of the signal is given by $\varrho_{t} =
1/p_2(t) \hbox{Tr}_2\left[\boldsymbol{R}\, \mathbb{I}\otimes \boldsymbol{\Pi}_{t}\right]$
where $p_2(t) = \sum_{s} p_{12}(s,t)$ is the marginal probability of measuring $t$
photoelectrons on the idler beam. After some calculations we arrive at
$$\varrho_{t} = \sum_{\boldsymbol q}
w_{t}(\gamma)\,
\theta(\gamma - t)
\:\bigotimes_{k=1}^\mu\: \ket{q_k}\bra{q_k}\:,$$
where $\theta(x)$ is the Heaviside step function,
$$w_{t}(\gamma)=
\binom{\gamma}{t}
\frac{\eta^t
(M_{t}-t\eta)^\gamma}
{(M_{t}+\mu\eta)^\gamma
[
p_2(t)
\left(1+ M /\eta\mu \right)^\mu
\left(1-\eta\right)^{t}
]}$$
and $$M_t=\hbox{Tr}_1[\varrho_t\,\sum_{s} s \boldsymbol{\Pi}_{s}] = [t
(M+\eta\mu)+ \mu M (1-\eta)] (M+\mu)^{-1}$$ is the mean number of
photoelectrons for the conditional state $\varrho_t$.  Similarly, upon
selecting a set of possible results $t\in {\cal T}$ according to a given
rule ${\cal T}$, a suitable engineering of the conditional state
$\varrho_{\cal T}$ may be achieved. As an example, we will consider the
states $$\varrho_{*}^{(\pm)}=\sum_{t \gtrless t^*} p_2(t)\,
\varrho_{t}$$
obtained by keeping the photoelectrons on the idler that are larger or
smaller than a given threshold $t^*$.
\par
The experimental setup is sketched in Fig.~\ref{setup}. The light source was a Nd:YLF
ps-pulsed laser (High-Q Laser Production, Austria) with built-in second and third
harmonic generation. The output at the fundamental ($1047$~nm) and that at the third
harmonics ($349$~nm) were used to produce a UV pump field ($261.75$~nm) via non-collinear
sum-frequency generation in a BBO crystal ($\beta$-BaB${}_2$O${}_4$, Castech, China, cut
angle $37^\circ$, $8$~mm length). The pump was then sent into another BBO crystal
(Kaston, China, cut angle $48.84^\circ$, $4$~mm length) to produce pairwise entanglement
at $523.5$~nm, in order to match the maximum quantum efficiency ($\eta\sim50\%$) of our
hybrid photodetectors (R10467U-40, Hamamatsu, Japan). For their photon-number resolving
power, these detectors proved to be useful in reconstructing detected-photon
distributions~\cite{bjmo09,bondani:ASL}. Here we exploit their features to perform both
conclusive and inconclusive conditional measurements of the photon number.  The UV stray
light was cut-off by a filter and by two harmonic separators. Signal and idler were
selected by two pin-holes (200 or 300~$\mu$m diameter, located at $1$~m from BBO2) in
order to minimize the number of collected modes. Notice that the number of temporal
modes, which is evaluated from the marginal detected-photon number distribution, cannot
be reduced at will. The only way to reduce the number of modes is to select a single
spatial mode, which involves the challenging matching of the collection areas in signal
and idler. The possible mismatch between the collection areas results in an effective
detection efficiency, reduced with respect to the nominal efficiency of the detectors,
which can be estimated through the level of noise reduction $R=\sigma^2(s-t)/\langle s+t
\rangle = 1-\eta$~\cite{ssh07} exhibited by two beams. For our beams we obtain, without
noise subtraction, $\eta \sim 0.06$. The light passing the pin-holes was coupled to two
multimode optical fibers and delivered to the detectors, whose outputs were amplified
(preamplifier A250 plus amplifier A275, Amptek), synchronously integrated (SGI, SR250,
Stanford), digitized (ATMIO-16E-1, National Instruments) and, finally, processed
off-line. Each experimental run was performed on $50\,000$ subsequent laser shots at
fixed values of the pump intensity.
\par
\begin{figure}[ht]
\includegraphics[width=.49\columnwidth]{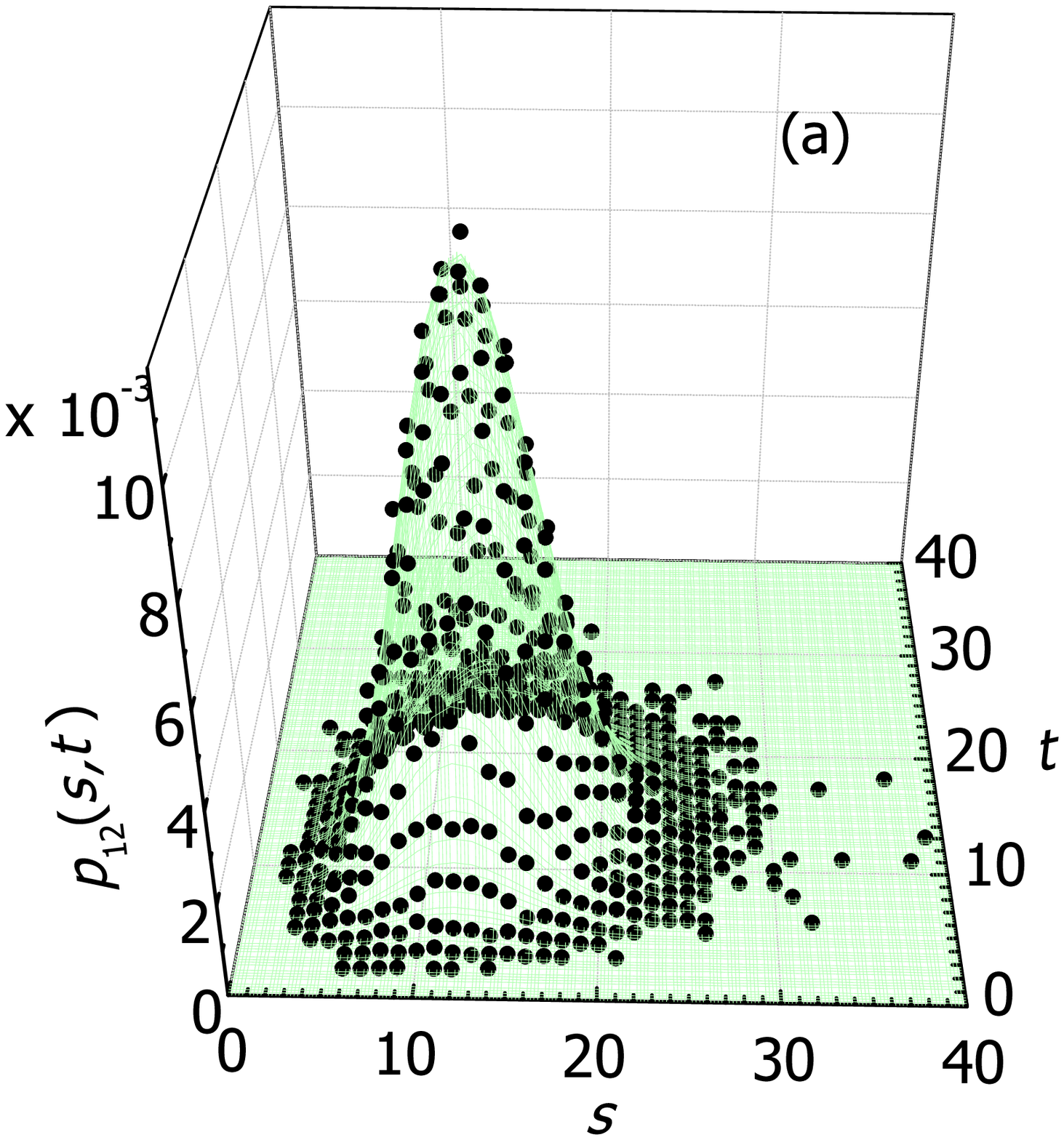}
\includegraphics[width=.49\columnwidth]{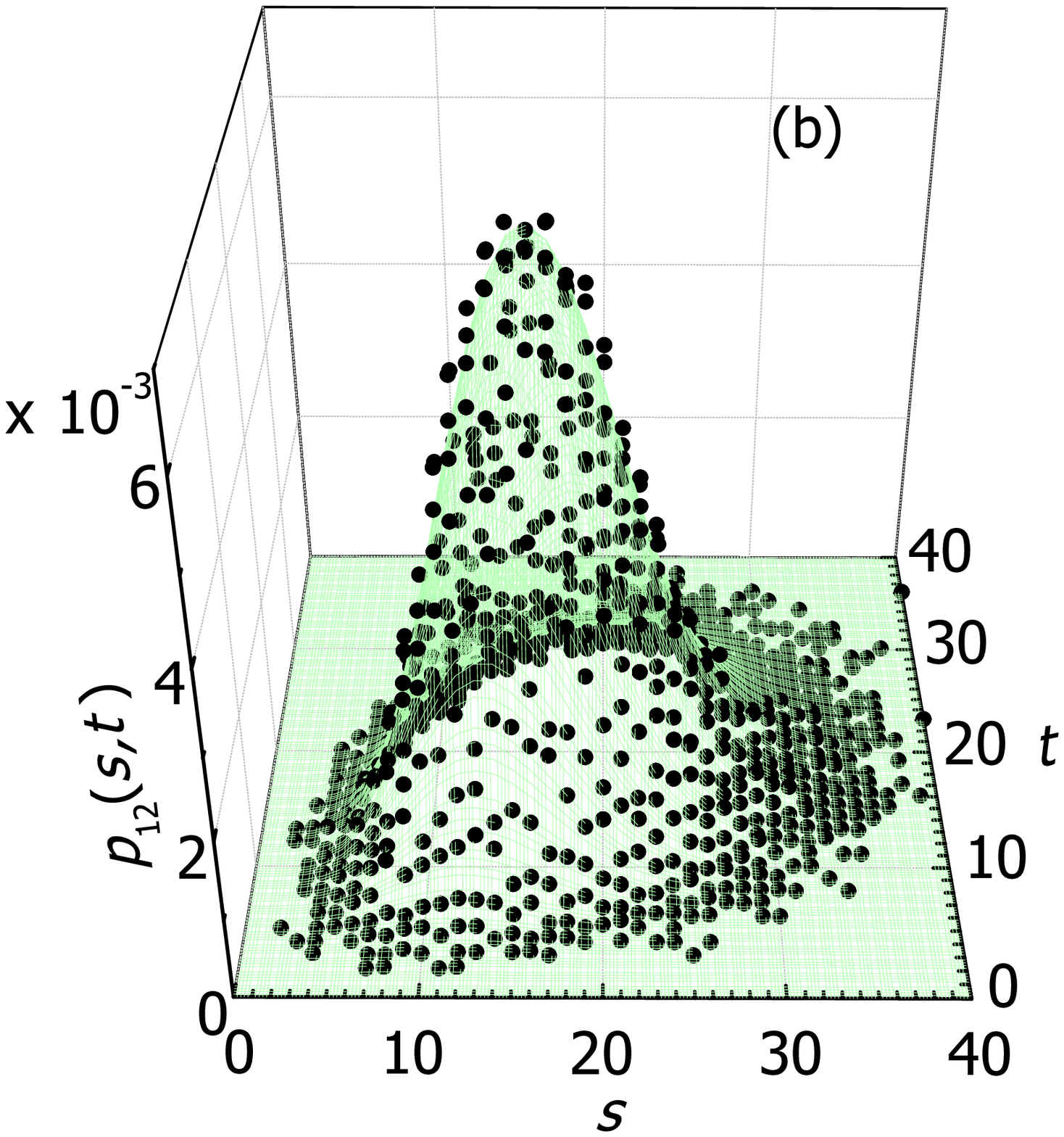}
\caption{(Color online) Joint probability distribution of photoelectrons
$p_{12}(s,t)$ compared to the experimental points. Panel (a):
$\mu = 197$, $\eta=0.06$ and $M=13.4$; panel (b):
$\mu = 25$, $\eta=0.056$ and $M=17.1$).}
\label{pjoin}
\end{figure}

As a first test of the correctness of our multimode description we checked the expression
of $p_{12}(s,t)$ against data: in Fig.~\ref{pjoin} we report the experimental joint
probability distribution superimposed to the theoretical one, evaluated for the
experimental values of the parameters (panel (a): PH~$= 200\mu$m, $\mu = 197$,
$\eta=0.06$ and $M=13.4$; panel (b): PH~=~300~$\mu$m, $\mu = 25$, $\eta=0.056$ and
$M=17.1$). The experimental results fit the theory very well and the fidelity
$\sum_{st}\sqrt{p_{12}^{th}(s,t)p_{12}^{exp}(s,t)}$ exceeds 0.99 for the whole range of
parameters. We also notice that the marginal probability distributions $p_1(s)$ and
$p_2(t)$, are multithermal distributions as it has been already observed in experiments
performed at different intensity regimes~\cite{pale04,ssh07,bjmo09}.
\par
\begin{figure}[ht]
\includegraphics[width=0.98\columnwidth]{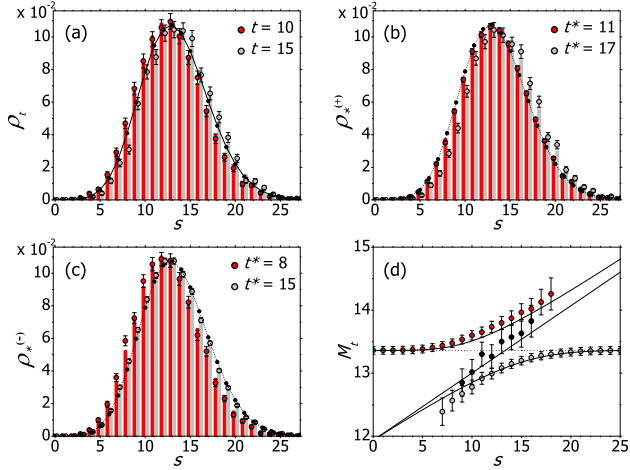}
\vspace{-0.7cm}
\caption{(Color Online) Photon distribution for conditional states.
(a): Experimental results (points) and theoretical distribution
(histograms) for the photoelectrons in the conditional signal state
$\varrho_{t}$ for $t=10$ (red histogram) and $t=15$ (gray histogram).
The black  line and the full circles represent respectively the
theoretical and experimental distribution for the unconditional state.
(b): As in panel (a) for $\varrho_{*}^{(+)}$, $t^*=11$ (red)
and $t^*=17$ (gray). (c): As in panel (a) for
$\varrho_{*}^{(-)}$, $t^*=8$ (red) and $t^*=15$ (gray).  Panel (d):
experimental mean value of the distributions as a function of the
conditioning value (or threshold).  Black circles refer to
$\varrho_{t}$, red circles to $\varrho_{*}^{(+)}$ and gray circles to
$\varrho_{*}^{(-)}$. Solid lines are the theoretical predictions
obtained for $M_t$.  The dashed line corresponds to the mean value of
the unconditioned state.  The other involved parameters are: $\mu =
197$, $\eta=0.06$ and $M=13.4$.} \label{condstates}
\end{figure}

In Fig.~\ref{condstates} we report the photon distributions $p_{1|2}(s|{\cal
T})=\hbox{Tr}_1[\varrho_{\cal T}\, \boldsymbol{\Pi}_{s}]$ of conditional states as
obtained from the state in Fig.~\ref{pjoin} (a) by choosing the values of the measured
photons on the idler beam according to a given rule and selecting the corresponding
ensemble on the signal beam. Panel~(a) of Fig.~\ref{condstates} displays the
distributions for the detected photons state $\varrho_{t}$ obtained by choosing a
definite number of detected photons ($t=10$ and $t=15$); panel~(b) those for
$\varrho_{*}^{(+)}$, obtained by keeping the values of detected photons larger than a
threshold $t^*$ ($t^*=11$ and $t^*=17$); finally, panel~(c) those for $\varrho_{*}^{(-)}$
($t^*=8$ and $t^*=15$). We notice that (i) the results are in excellent agreement with
theory and (ii) despite the small value of effective quantum efficiency the
``conditioning power'' of the measurement (\emph{i.e.} the differences between the
conditional states and the corresponding original ones) is appreciable. This is clearly
illustrated by the behavior of the mean values of the distributions, which are reported
in panel~(d) of Fig.~\ref{condstates} as a function of either the conditioning value or
the threshold: the experimental data are in excellent agreement with the predictions for
$M_t$.
\par
\begin{figure}[ht]
\includegraphics[width=0.98\columnwidth]{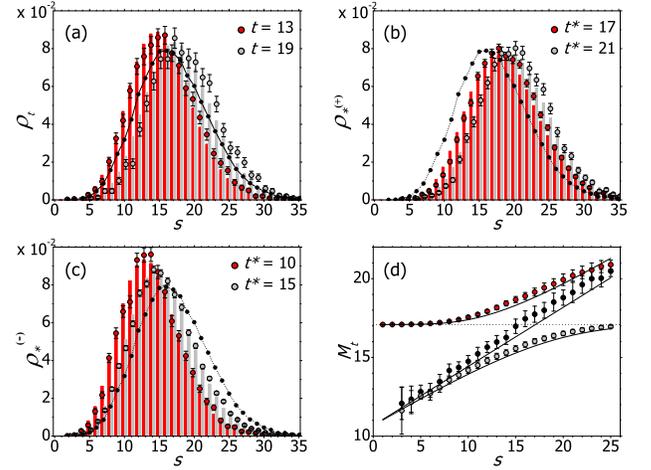}
\vspace{-0.7cm}
\caption{(Color Online) Photon distribution for conditional states.
(a): Experimental results (points) and theoretical distribution
(histograms) for the photoelectrons in the conditional signal state
$\varrho_{t}$ for $t=13$ (red histogram) and $t=19$ (gray histogram).
The black  line and the full circles represent respectively the
theoretical and experimental distribution for the unconditional state.
(b): As in panel (a) for $\varrho_{*}^{(+)}$, $t^*=17$ (red)
and $t^*=21$ (gray). (c): As in panel (a) for
$\varrho_{*}^{(-)}$, $t^*=10$ (red) and $t^*=15$ (gray).  Panel (d):
experimental mean value of the distributions as a function of the
conditioning value (or threshold).  Black circles refer to
$\varrho_{t}$, red circles to $\varrho_{*}^{(+)}$ and gray circles to
$\varrho_{*}^{(-)}$. Solid lines are the theoretical predictions
obtained for $M_t$.  The dashed line corresponds to the mean value of
the unconditioned state.  The other involved parameters are: $\mu =
25$, $\eta=0.056$ and $M=17.1$.} \label{condstatesLOW}
\end{figure}

In Fig.~\ref{condstatesLOW} we show the results for another dataset having similar mean
value and a considerably lower number of modes. We note that the results are again in
excellent agreement with theory and the efficacy of the ``conditioning power'' of the
measurement is more evident with respect to the case of a larger number of modes.
\par
\begin{figure}[ht]
\includegraphics[width=0.49\columnwidth]{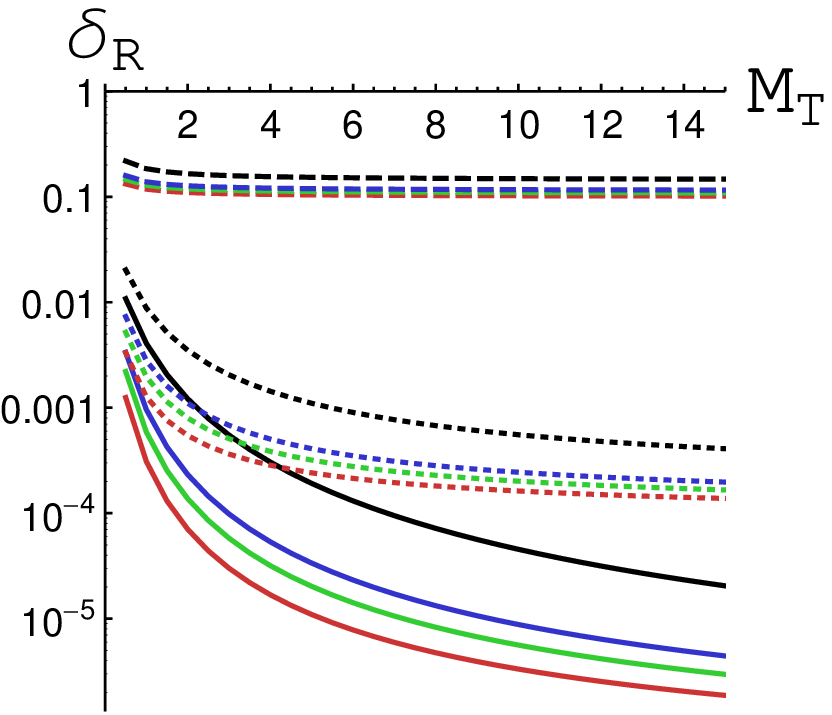}
\includegraphics[width=0.47\columnwidth]{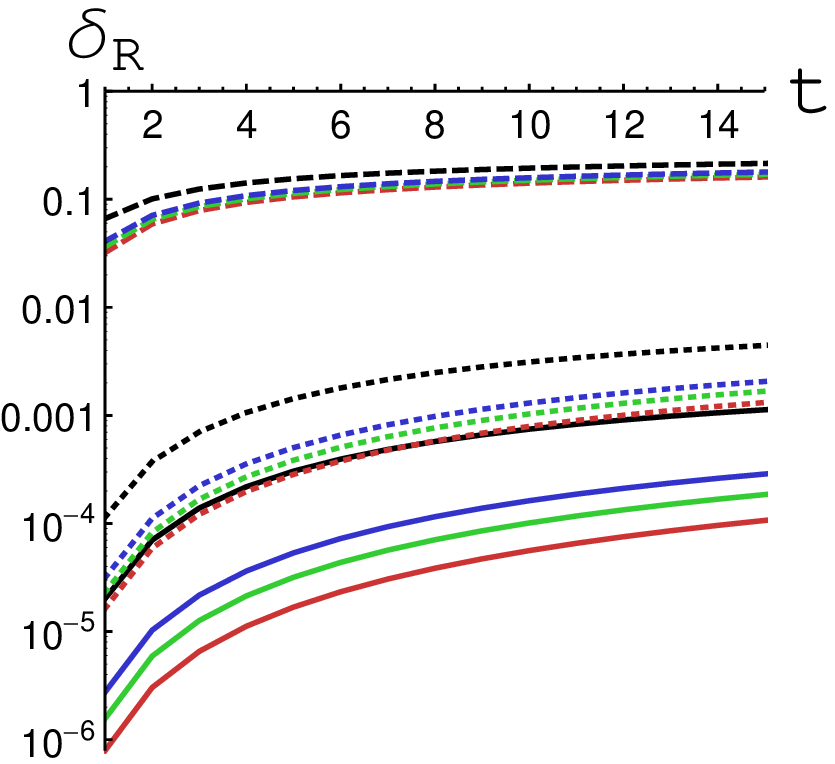}
\includegraphics[width=0.49\columnwidth]{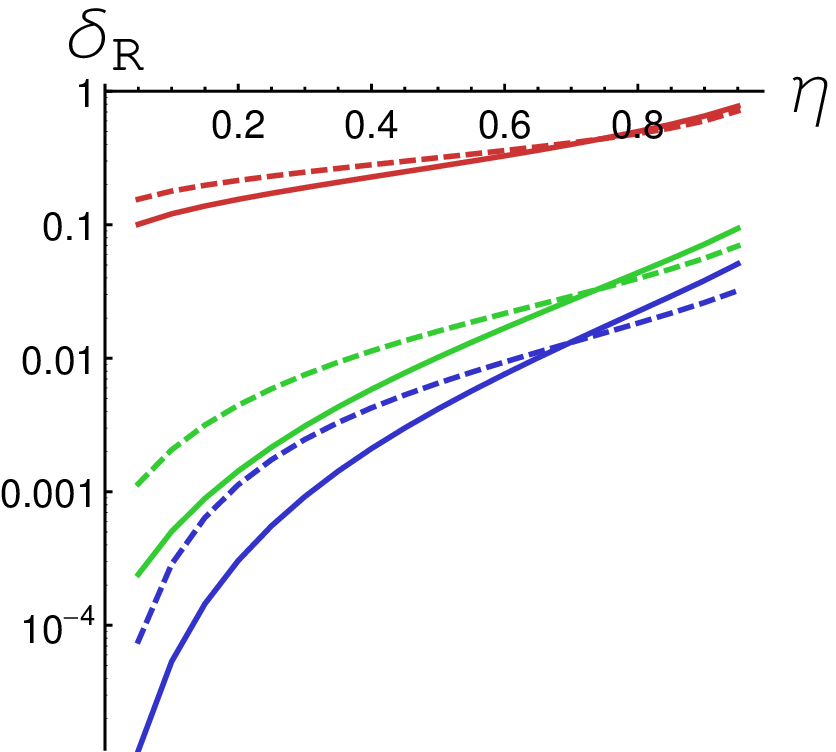}
\includegraphics[width=0.47\columnwidth]{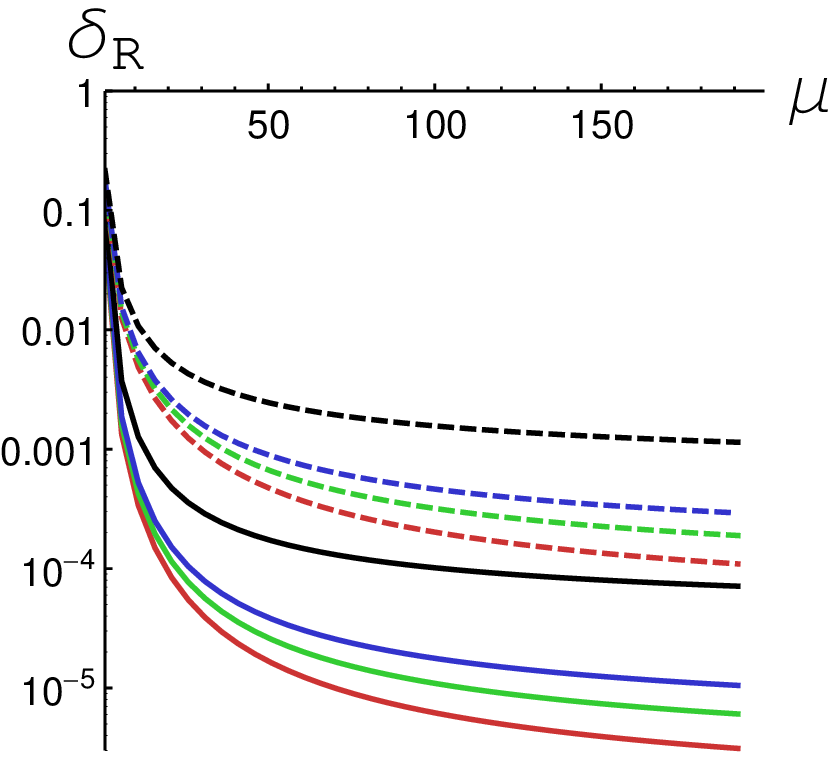}
\caption{(Color Online) Renormalized nonGaussianity $\delta_{\R}$
of the conditional state $\varrho_t$ for different values of the
experimental parameters.
(\underline{Top left panel}):
$\delta_{\R}$ as a function of the total number of measured
photoelectrons $M_t$ for $t=5$ and for different values of the quantum
efficiency $\eta$ and of the number of modes $\mu$. Solid lines are for
$\mu=197$, dotted for $\mu=25$ and dashed for $\mu=1$. For each group of
lines we have, from bottom to top, $\eta=6\%$ (red), $\eta=8\%$ (green),
$\eta=10\%$ (blue) and $\eta=20\%$ (black).
(\underline{Top right panel}):
$\delta_{\R}$ as a function of the conditioning number $t$ of detected
photoelectrons on the idler beam for $M_t=4$ and for different values of
the quantum efficiency $\eta$ and of the number of modes $\mu$. Solid lines
are for $\mu=197$, dotted for $\mu=25$ and dashed for $\mu=1$. For each
group of lines we have, from bottom to top, $\eta=6\%$ (red), $\eta=8\%$
(green), $\eta=10\%$ (blue) and $\eta=20\%$ (black).
(\underline{Bottom left panel}):
$\delta_{\R}$ as a function of the quantum efficiency $\eta$ for $M_t=4$
and for different values of the conditioning number $t$ of detected
photoelectrons and of the number of modes $\mu$.  Solid lines are for
$t=5$ and dashed ones for $t=15$.  From bottom to top we can see lines
for $\mu=197$ (blue), $\mu=25$ (green), and $\mu=1$ (red).
(\underline{Bottom right panel}):
$\delta_{\R}$ as a function of the number of modes $\mu$ for $M_t=4$ and
for different values of conditioning number $t$ of detected photoelectrons
and for the quantum efficiency $\eta$.  Solid lines are for
$t=2$ and dashed for $t=15$. For each group of lines we have, from
bottom to top, $\eta=6\%$ (red), $\eta=8\%$ (green), $\eta=10\%$ (blue)
and $\eta=20\%$ (black).
} \label{f:ng}
\end{figure}

The nonGaussian character of the conditional states may be foreseen from the deviation of
the detected-photon statistics from that of the original state. However, the shape of the
distributions (not too different from the unconditioned ones) and the low value of the
quantum efficiency anticipate that the amount of nonGaussianity will be unavoidably
small. In order to assess the performances of our scheme we focus on the conditional
state $\varrho_{t}$ and use the nonGaussianity measure
$\delta[\varrho]=S[\tau]-S[\varrho]$, where $S[\varrho]$ is the Von Neumann entropy of
the state $\varrho$ and $\tau$ is the Gaussian reference state of $\varrho$, \emph{i.e.},
a Gaussian state with the same mean value and covariance matrix as $\varrho$. $\delta$
has been proved to be a proper measure of nonGaussianity\cite{genoni:nonG}, as well as a
critical parameter to asses nonGaussianity as a resource \cite{EKerr}. In our case $\tau$
is a factorized thermal state with $M_{t}/\eta\mu$ mean photons per
mode~\cite{genoni:nonG} and the Von Neumann entropy of the conditional state is given by
$S[\varrho_{t}] = -\sum_{\gamma=0}^\infty \binom{\gamma+\mu-1}{\gamma} w_{t}(\gamma) \log
w_{t}(\gamma)$. By using the above expression and the Von Neumann entropy of a factorized
thermal state, we evaluate the nonGaussianity and normalize its value to that of a
maximally nonGaussian state for the same mean number of photons and modes, i.e., a
factorized Fock state \cite{genoni:nonG}. The renormalized nonGaussianity
$\delta_{\R}[\varrho_t]$ is reported in Fig. \ref{f:ng} for different values of the
experimental parameters.
\par
As it is apparent from the plots the renormalized nonGaussianity $\delta_{\R}$ is a
decreasing function of the energy of the conditional state and of the number of modes.
The effect of the quantum efficiency is more relevant for large number of modes and large
energy. As it concerns the conditioning value $t$ of detected photons, we see that
$\delta_{\R}$ monotonically increases with $t$, again with the quantum efficiency playing
a major role for large number of modes. Overall, the goal of achieving high
nonGaussianity requires a small number of modes or, at fixed number of modes, a high
value of the quantum efficiency. Since the pulsed nature of the PDC pump unavoidably
leads to a multimode output, the performances of the present source in the generation of
nonGaussian states may be improved by increasing the overall quantum efficiency,
\emph{i.e.} the matching of the collection areas in signal and idler beams.
\par
In conclusion, we have suggested and demonstrated a novel bright source of
\textbf{nonclassical} states based on multimode spontaneous PDC and conditional intensity
measurements. We have developed a theoretical model based only on experimentally
accessible quantities and derived closed formulas for the detection probability, the
conditional states and the corresponding nonGaussianity.  We have compared our
predictions with experimental data and found an excellent agreement in the whole range of
accessible experimental parameters. Our results clearly indicate the possibility of
quantum state engineering with multiphoton/multimode conditional states using mesoscopic
photon counting and multimode pairwise correlated states.
\par
This work has been partially supported by the CNR-CNISM agreement.


\begin{thebibliography}{99}
\bibitem{ngOPO1} V.~D'Auria, C.~de Lisio, A.~Porzio, S.~Solimeno, J.~Anwar and M.~G.~A.~
    Paris, \emph{Phys. Rev. A}, {\bf 81} (2010) 033846.
\bibitem{ngOPO2} A.~Chiummo, M.~De Laurentis, A.~Porzio, S.~Solimeno and M.~G.~A.~Paris,
    \emph{Opt. Expr.}, {\bf 13} (2005) 948.
\bibitem{kerr:PRL:01} C.~Silberhorn, P.~K.~Lam, O.~Wei\ss{}, F.~K\"{o}nig, N.~Korolkova,
    and G.~Leuchs, \emph{Phys. Rev. Lett.}, {\bf 86} (2001) 4267.
\bibitem{kerr:PRA:06} O.~Gl\"{o}ckl, U.~L.~Andersen and G.~Leuchs, \emph{Phys. Rev. A}
    {\bf 73} (2006) 012306.
\bibitem{kerr:NJP:08} T.~Tyc and N.~Korolkova, \emph{New J. Phys.}, {\bf 10},
    (2008) 023041.
\bibitem{eng} M.~G.~A.~Paris, M.~Cola and R.~Bonifacio, \emph{Phys. Rev. A}, {\bf 67},
    (2003) 042104.
\bibitem{dar} G.~M.~D'Ariano, P.~Kumar, C.~Macchiavello, L.~Maccone and N.~Sterpi,
    \emph{Phys. Rev. Lett.}, {\bf 83} (1999) 2490.
\bibitem{wel} T. Opatrn\'y, G. Kurizki and D. -G. Welsch, \emph{Phys. Rev. A}, {\bf 61},
    (2000) 032302; M. Dakna, L. Kn\"{o}ll and D. -G. Welsch,
    \emph{Opt. Comm.}, {\bf 145} (1998) 309.
\bibitem{coc} P.~T.~Cochrane, T.~C.~Ralph and G.~J.~Milburn, \emph{Phys. Rev. A}, {\bf
    65} (2002) 062306.
\bibitem{oli} S.~Olivares, M.~G.~A.~Paris and R.~Bonifacio, \emph{Phys. Rev. A}, {\bf
    67} (2003) 032314; S.~Olivares, M.~G.~A.~Paris, \emph{J. Opt. B}, {\bf
    7} (2005) S392.
\bibitem{oli1} S.~Olivares and M.~G.~A. Paris, \emph{J. Opt. B}, {\bf 7} (2005) 616; C.~
    Invernizzi, S.~Olivares, M.~G.~A.~Paris and K.~Banaszek, \emph{Phys. Rev. A}, {\bf
    72} (2005) 042105.
\bibitem{mar:PRA:10} P.~Marek and R.~Filip, \emph{Phys. Rev. A}, {\bf 81} (2010) 022302.
\bibitem{all10} A. Allevi, A. Andreoni, M. Bondani, M. G. Genoni, S. Olivares, 
\emph{Phys. Rev. A} (2010), in press.
\bibitem{pgb} D.~T.~Pegg, L.~S.~Philips and S.~M.~Barnett, \emph{Phys. Rev. Lett.}, {\bf
    81} (1998) 1604.
\bibitem{cat1} M.~Dakna, T.~Anhut, T.~Opatrny, L.~Kn\"{o}ll and D.~G.~Welsch,
    \emph{Phys. Rev. A}, {\bf 55} (1997) 3184.
\bibitem{cat2} M.~Dakna, J.~Clausen, L.~Kn\"{o}ll and D.-G.~Welsch,
    \emph{Acta Phys. Slov.}, {\bf 48} (1998) 207.
\bibitem{cat3} S.~B.~Zheng, \emph{Phys. Lett. A}, {\bf 245} (1998) 11.
\bibitem{ff} G.~M.~D'Ariano, L.~Maccone, M.~G.~A.~Paris and M.~F.~Sacchi, \emph{Phys.
    Rev. A}, {\bf 61} (2000) 053817; \emph{Fort. Phys.}, {\bf 48} (2000) 511.
\bibitem{zlc} Lu-Ming Duan, G.~Giedke, J.~I.~Cirac and P.~Zoller, \emph{Phys. Rev.
    Lett.}, {\bf 84} (2000) 4002.
\bibitem{uni} B.~Hladky, G.~Drobny and V.~Buzek, \emph{Phys. Rev. A}, {\bf 61},
    (2000) 022102.
\bibitem{opq} M.~G.~A.~Paris, \emph{Phys. Rev. A}, {\bf 62} (2000) 033813.
\bibitem{cla} J.~Clausen, M.~Dakna, L.~Kn\"{o}ll and D.-G.~Welsch
    \emph{J. Opt. B}, {\bf 1} (1999) 332; M.~G.~A.~Paris, Phys. Lett. A {\bf 217} (1996) 78.
\bibitem{sab} A.~Napoli, A.~Messina and S.~Maniscalco, \emph{Acta Phys. Slov.}, {\bf 50},
    (2000) 519.
\bibitem{pla} F.~Plastina and F.~Piperno, \emph{Eur. Phys. J. D}, {\bf 5} (1999) 411.
\bibitem{bno} M.~Ban, \emph{Opt. Comm.}, {\bf 143} (1997) 225.
\bibitem{koz} A.~Kozhekin, G.~Kurizky and B.~Sherman, \emph{Phys. Rev. A}, {\bf 54},
    (1996) 3535.
\bibitem{kim} M.~S.~Kim, \emph{J. Phys. B}, {\bf 41} (2008) 133001.
\bibitem{eisert:entdist} J.~Eisert, S.~Scheel and M.~B.~Plenio, \emph{Phys. Rev. Lett.}
    {\bf 89} (2002) 137903.
\bibitem{fiur:entdist} J.~Fiur\'a\ifmmode~\check{s}\else \v{s}\fi{}ek, \emph{Phys. Rev.
    Lett.}, {\bf 89} (2002) 137904.
\bibitem{ngnp1} R.~Dong, M.~Lassen, J.~Heersink, C.~Marquardt, R.~Filip, G.~Leuchs and
    U.~L.~Andersen, \emph{Nature Phys.}, {\bf 4} (2008) 919.
\bibitem{ngnp2} T.~Aoki, G.~Takahashi, T.~Kajiya, J.~Yoshikawa, S.~L.~Braunstein, P.~van
    Loock and A.~Furusawa, \emph{Nature Phys.}, {\bf 5} (2009) 541.
\bibitem{inv:PRA:05} C.~Invernizzi, S.~Olivares, M.~G.~A.~Paris and K.~Banaszek,
    \emph{Phys. Rev. A}, {\bf 72} (2005) 042105.
\bibitem{c1} N.~J.~Cerf, O.~Kr\"{u}ger, P.~Navez, R.~F.~Werner and M.~M.~Wolf,
    \emph{Phys. Rev. Lett.}, {\bf 95} (2005) 070501.
\bibitem{cs07} F. Casagrande, A. Lulli and M. G. A. Paris, \emph{Phys. Rev. A}, {\bf 75},
    (2007) 032336.
\bibitem{ourjo:entdist} A.~Ourjoumtsev, A.~Dantan, R.~Tualle-Brouri and P.~Grangier,
    \emph{Phys. Rev. Lett.}, {\bf 98} (2007) 030502.
\bibitem{taka09} H.~Takahashi, J.~S.~Neergaard-Nielsen, M.~Takeuchi, M.~Takeoka,
    K.~Hayasaka, A.~Furusawa and M.~Sasaki, \emph{Nature Phot.}, {\bf 4} (2010) 178.
\bibitem{gerrits10} T.~Gerrits, S.~Glancy, T.~S.~Clement, B.~Calkins, A.~E.~Lita,
    A.~J.~Miller, A.~L.~Migdall, S.~W.~Nam, R.~P.~Mirin and E.~Knill, arXive:1004.2727v1
\bibitem{mas06} M.~Sasaki and S.~Suzuki, \emph{Phys. Rev. A}, {\bf 73} (2006) 043807.
\bibitem{ps10} M.~G.~Genoni, F.~A.~Beduini, A.~Allevi, M.~Bondani, S.~Olivares and
    M.~G.~A.~Paris, \emph{Phys. Scripta}, {\bf T139} (2010) in press
\bibitem{jd05} A.~Agliati, M.~Bondani, A.~Andreoni, G.~De~Cillis and M.~G.~A.~Paris,
    \emph{J. Opt. B}, {\bf 7},( 2005) 652.
\bibitem{Was06} W. Wasilewski, A. I. Lvovsky, K. Banaszek, C. Radzewicz.
 \emph{Phys. Rev. A} {\bf 73}, (2006) 063819.
\bibitem{ssh07} M.~Bondani, A.~Allevi, G.~Zambra, M.~G.~A.~Paris and A.~Andreoni,
    \emph{Phys. Rev. A}, {\bf 76} (2007) 013833 .
\bibitem{Lvo07} A. I. Lvovsky, W. Wasilewski, K Banaszek, 
\emph{J. Mod. Opt} {\bf 54}, (2007), 721.
\bibitem{Was08} W. Wasilewski, C. Radzewicz, R. Frankowski, K. Banaszek,
    \emph{Phys. Rev. A}, {\bf 78} (2008) 033831.
\bibitem{sil09a} W.~Mauerer, M.~Avenhaus, W.~Helwig and C.~Silberhorn, \emph{Phys. Rev.
    A}, {\bf 80} (2009) 053815.
\bibitem{sil09b} W.~Helwig, W.~Mauerer and C.~Silberhorn \emph{Phys. Rev. A} {\bf 80},
    (2009) 052326.
\bibitem{sil10} C.~S\"{o}ller, B.~Brecht, P.~J.~Mosley, L.~Y.~Zang, A.~Podlipensky,
    N.~Y.~Joly, P.~St.~J.~Russell and C.~Silberhorn, \emph{Phys. Rev. A}, {\bf 81},
    (2010), 031801.
\bibitem{bjmo09} M.~Bondani, A.~Allevi, A.~Agliati and A.~Andreoni, \emph{J. Mod. Opt.}
    {\bf 56} (2009) 226.
\bibitem{pale04} F.~Paleari, A.~Andreoni, G.~Zambra and M.~Bondani, \emph{Opt. Express}
    {\bf 12} (2004) 2816.
\bibitem{bondani:ASL} M.~Bondani, A.~Allevi and A.~Andreoni, \emph{Adv. Sci. Lett.}, {\bf
    2} (2009) 463;\emph{ Opt. Letters}, {\bf 34} (2009) 1444.
\bibitem{genoni:nonG} M.~G.~Genoni, M.~G.~A.~Paris and K.~Banaszek, \emph{Phys. Rev. A}
    {\bf 76} (2007) 042327; \emph{Phys. Rev. A}, {\bf 78} (2008) 060303(R).
\bibitem{EKerr} M.~G.~Genoni, C.~Invernizzi and M.~G.~A.~Paris, \emph{Phys. Rev. A}, {\bf
    80} (2009) 033842.
\end{thebibliography}
\end{document}